# Signal Peak-Tracker based on the Teager-Kaiser Energy (TKE) Operator

Randall D. Peters

Physics Department
Mercer University, Macon, GA

**Abstract**

Described is a modification of the TKE operator from its usual `energy form'. The resulting `peak-tracker' (or peak-detector) is especially useful in studies that involve the frequency domain.

**Background**

The Teager-Kaiser Energy (TKE) operator originated through (i) considerations of speech nonlinearity and (ii) the physics of the simple harmonic oscillator. References to its historical development, along with extensive treatments of its attributes, are provided in Ref. [1]. In applications familiar to the present author, the primary function of TKE is to operate as a demodulaor, of both amplitude modulated (AM) and frequency modulated (FM) signals. Though it can accommodate some FM signal cases, it is more effective in the (near) real-time extraction of information content from signals of AM-type.

In studies involving the frequency domain, it is more convenient to work with a peak-detector than with an energy detector. A direct quantitative comparison can then be made of demodulation spectral lines obtained using (i) the old-fashion-method of rectification, and (ii) the present method, labeled here as Teager-Kaiser-Peak (TKP). The modification of TKE yielding the TKP operator is next described; it is one that yields a properly-normalized peak-detector.

**Simple Harmonic Oscillator (SHO)**

The SHO is one of the most important `foundation-stones' of theoretical physics. Its equation of motion (in non-driven Hooke's-law form) is given by:

$$\ddot{x} + \frac{\omega_o}{Q}\dot{x} + \omega_o^2 x = 0 \qquad (1)$$

where Q is the quality factor, related to the (viscous exponential) damping parameter β through $Q = \omega_o/(2\beta)$.

For the Hooke's law case of a mass/spring, the displacement of the mass from equilibrium is specified by x, and the mass value is implicit in the equation through the natural (angular) frequency term according to

$$\omega_o^2 = \frac{k}{m} \qquad (2)$$

where k is the spring constant.

The total energy of the system is obtained by adding the kinetic energy of the mass ( m (dx/dt)$^2$/2) and the potential energy of the spring (k x$^2$/2), which is expressible as

$$\frac{E}{m} = \frac{1}{2}(\dot{x}^2 + \omega_o^2 x^2) \qquad (3)$$

Eq. 1 is readily integrated numerically, using Microsoft's Excel spreadsheet, by working with its equivalent pair of coupled, discrete first order equations; i.e.,

$$v_n = v_{n-1} - \left(\frac{\omega_o}{Q} v_{n-1} + \omega_o^2 x_{n-1}\right)\delta t \quad , \quad x_n = x_{n-1} + v_n \delta t \qquad (4)$$

where the time-step between updates ($\delta t = 1/F_s$) and $F_s$ is the sample rate. During the integration, the v-update of each iteration must precede the x-update (last-point-approximation as opposed to the unstable first-point-approximation, also known as the Euler algorithm). These were performed with the following conditions on the parameters:

All values of the state variables (displacement x, velocity v = dx/dt, and acceleration, a = dv/dt = d$^2$x/dt$^2$) were set initially to 0 except for x (t = 0) = 1 m, Q = 10, $\omega_o$ = 3.1416 s$^{-1}$, and $\delta t$ = 1 ms.

Results from one example case that was selected are shown in Fig. 1. To illustrate the `tracking' features of the TKE and (the later described) TKP operators, the sign of Q was reversed halfway through the 10 s time interval of the simulation.

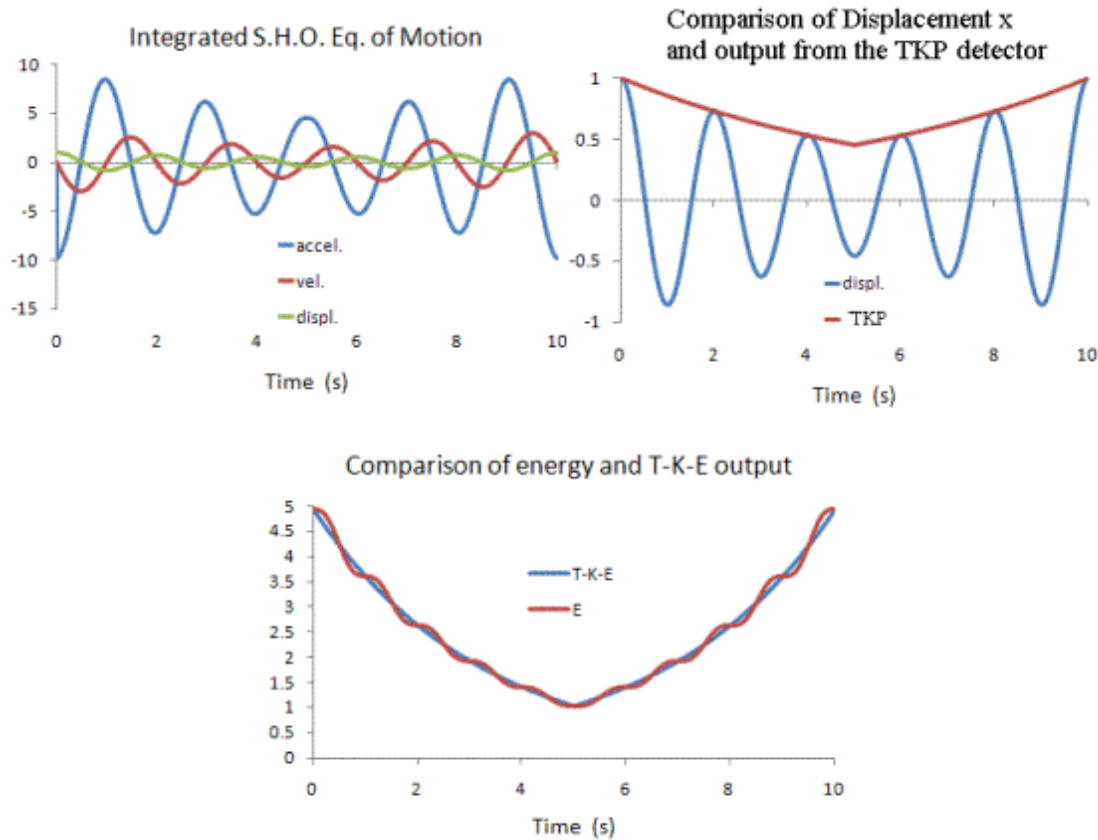

**Figure 1.** Results of a numerical integration of the SHO equation of motion. In the first half of the 10 s interval shown, the oscillator decayed exponentially. Thereafter, in the 2nd half of the interval, it was forced (through simulated negative damping) to rise exponentially back to the initial state.

The exponential decay, thereafter followed by the exponential rise of the turning points of each of the state variables, is clearly visible in the upper left graph of Fig. 1. In the bottom graph are plots of (i) the energy E/m according to Eq. 3 (red) and (ii) the estimate of the energy, based on the following continuous form of the TKE operator (blue)

$$E_{TKE} = \frac{1}{2}(\dot{x}^2 - x \cdot \ddot{x}) \qquad (5)$$

In most examples from literature, the factor of 1/2 shown in Eq. (5) is not included. For quantitative work, as opposed to looking at relative changes, the factor is required. Note also, that the TKE operator does not provide an estimate that is in exact agreement with the actual total (specific) energy. Rather, it agrees with a per-cycle average of the energy, as seen from the figure. It is not widely-enough appreciated that the viscous damped oscillator total energy is not a pure exponential decay. Rather, it always exhibits the cyclic variability around exponential that can be seen in the figure. The cyclic difference between the two energies is due to a term that is missing from Eq.(5). An exact form for the total energy must include work done by the friction force per unit mass, i.e.,

$$E_{exact} = \frac{1}{2}[\dot{x}^2 - x \cdot (\ddot{x} + \frac{\omega_o}{Q}\dot{x})] \qquad (6)$$

Although Eq. 6 can be shown to be in perfect agreement with the total energy, its implementation and use instead of Eq.(5) is probably not practical in the majority of applications.

**Discrete Form of the TK operator**

In Ref.[1] the following discrete form of the Teager operator is derived from the continuous form Eq.(5).

$$\Psi[x_n] = x_n^2 - x_{n-1}x_{n+1} \qquad (7)$$

It is further shown there that $\Psi[x_n] = A^2\Omega^2$ where $\Omega = \omega_o/F_s$, with $F_s$ being the sample rate, if small enough that $\sin\Omega \approx \Omega$. With that derivation one obtains the following peak detector operator:

$$TKP = [(F_s/\omega_o)^2 (x_n^2 - x_{n-1} x_{n+1})]^{1/2} \qquad (8)$$

Eq.(8) was used in the Excel code that generated the curves of Fig. 1 and it is seen from the upper-right graph (red) to be an excellent predictor for the peak values of the time varying displacement (blue). In considering the use of Eq. 8, one should be careful to check that $F_s/\omega_o \gg 1$. If this condition is not met, then the term should be replaced by $1/\sin(\omega_o/F_s)$.

As a further test of the peak tracker, another simulation was considered for this study, the results of which are shown in Fig. 2. For this case, a series of contiguous exponential decays was defined, where nearest neighbors of the five parts differ in amplitude from each other by a significant amount. In other words, a sharp change in amplitude rate occurs at the transition points.

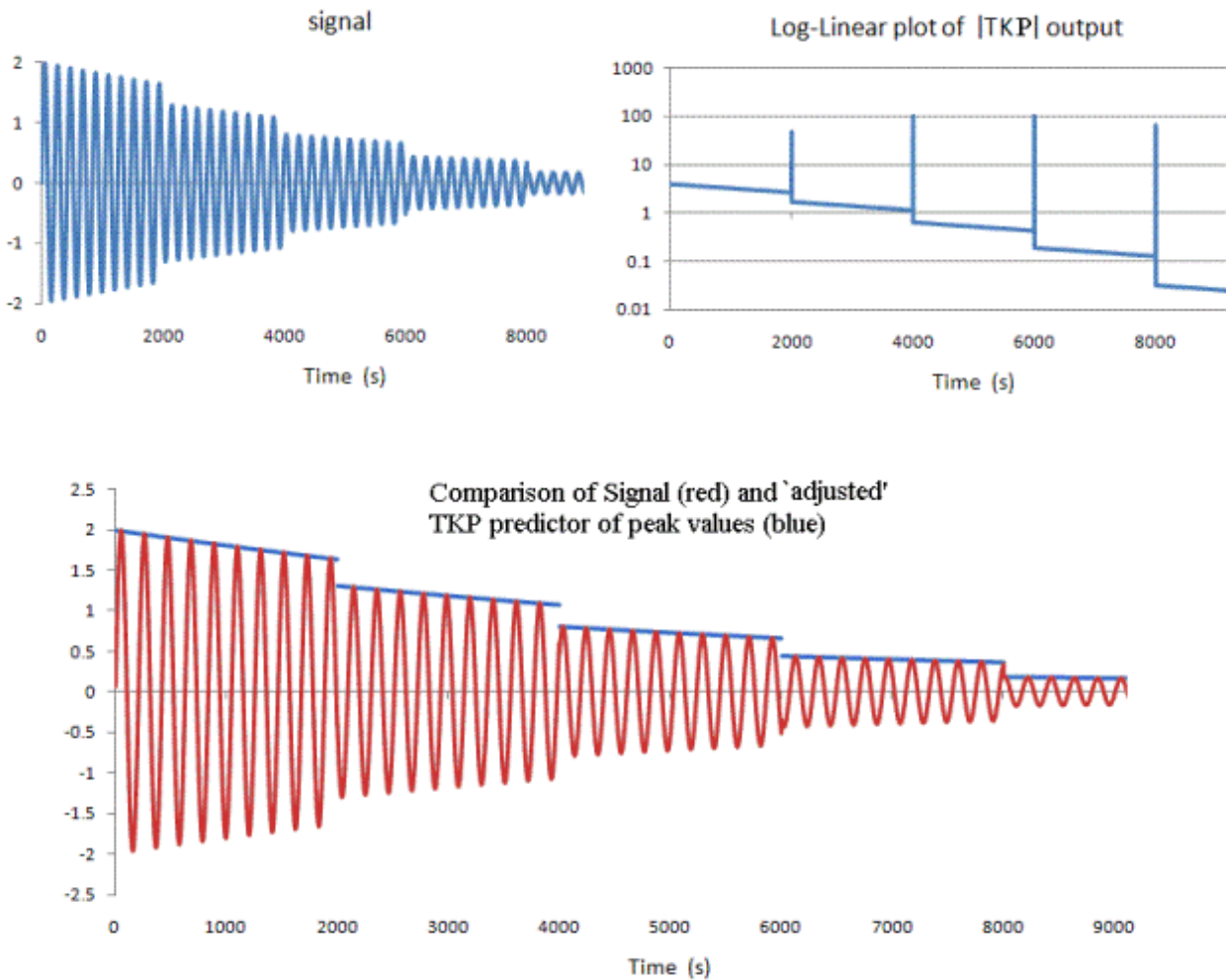

**Figure 2.** Simulation (top left) used to generate TKP detector predictions (top right) using Eq.8. Anomalies associated with abrupt changes in the peak have been removed in the bottom graph.

Where the changes are nearly discontinuous, the operator yields large anomalous values that can be both positive and negative. For the top right graph, only absolute values were plotted, to avoid imaginary results from the square-root calculation and so that a log-ordinate could be used to show the size of the anomalies. Involving only a pair of points either side of the transition, these `out-lying' values were culled from the bottom graph. It is seen from this bottom graph that the `adjusted' set faithfully tracks all parts of the time-changing amplitudes, just as was also true for the case illustrated by Fig.1.